\documentclass{article}
\usepackage[T1]{fontenc} 
\usepackage[utf8]{inputenc} 
\usepackage{ismir,amsmath,cite,url,multirow,makecell, balance}
\usepackage{graphicx}
\usepackage{color}

\usepackage{lineno}

\title{Improving Real-Time Music Accompaniment Separation with MMDenseNet}

\multauthor
{Chun-Hsiang Wang$^1$ \hspace{1cm} Chung-Che Wang$^1$ \hspace{1cm} Jun-You Wang$^1$} { \bfseries{Jyh-Shing Roger Jang$^1$ \hspace{1cm} Yen-Hsun Chu$^2$ }\\
 $^1$ Dept. of Computer Science and Information Engineering, National Taiwan Univ., Taiwan\\
$^2$ Realtek Semiconductor Corp.\\
}

\def\authorname{F. Author, S. Author, and T. Author}

\begin{document}

\maketitle
\begin{abstract}
Music source separation aims to separate polyphonic music into different types of sources. Most existing methods focus on enhancing the quality of separated results by using a larger model structure, rendering them unsuitable for deployment on edge devices. Moreover, these methods may produce low-quality output when the input duration is short, making them impractical for real-time applications. Therefore, the goal of this paper is to enhance a lightweight model, MMDenstNet, to strike a balance between separation quality and latency for real-time applications. Different directions of improvement are explored or proposed in this paper, including complex ideal ratio mask, self-attention, band-merge-split method, and feature look back. Source-to-distortion ratio, real-time factor, and optimal latency are employed to evaluate the performance. To align with our application requirements, the evaluation process in this paper focuses on the separation performance of the accompaniment part. Experimental results demonstrate that our improvement achieves low real-time factor and optimal latency while maintaining acceptable separation quality.
\end{abstract}
\section{Introduction}
\label{sec:intro}

Music source separation aims to separate polyphonic music into different types of sources, such as vocal, drum, piano, or other instruments. Singing voice separation is not only a research topic in itself, but good separation results can also benefit downstream applications, including voice conversion~\cite{rajpura2020effectiveness}, lyrics alignment~\cite{gao2022music}, and accompaniment for karaoke~\cite{liu2020voice}. In this paper, we aim to extract the accompaniment part in real-time with low latency for a karaoke application.

With the rapid development of deep learning, the effectiveness of singing voice separation has significantly improved. HT Demucs~\cite{rouard2023hybrid} is the state-of-the-art model (among those with official implementation before mid-2024), which employs two U-Nets that respectively process waveform and spectrogram. It combines the information from both U-Nets using a Transformer and cross-attention at the bottleneck layer. Despite the high separation quality of HT Demucs, its larger number of parameters and high latency on CPU make it challenging for real-time applications on edge devices. On the other hand, the Multi-scale multi-band DenseNet (MMDenseNet)~\cite{takahashi2017multi} is one of the relatively lightweight source separation models. Despite MMDenseNet's slightly lower separation quality, its fast speed makes it more suitable for real-time applications.

To further reduce the latency while still maintaining comparable separation quality, we propose or invoke various methods for improving the raw version of MMDenseNet, including the complex ideal ratio mask (cIRM)~\cite{kong2021decoupling}, self-attention, the band-merge-split method inspired by~\cite{luo2023music}, and feature look back. Note that the above methods are possible for benefiting other separation model structures like band-split RNN~\cite{luo2023music}, U-Net~\cite{jansson2017singing}, Open-Unmix~\cite{stoter2019open}, and KUIELab-MDX-Net~\cite{kim2021kuielab}. We choose MMDenseNet for this paper because our preliminary experiments show that MMDenseNet achieves the best trade-off for both quality and speed for our application requirements.

The rest of this paper is organized as follows. Section~\ref{sec:method} describes our methods for improving MMDenstNet, Section~\ref{sec:exp} shows the experimental results, and Section~\ref{sec:con} concludes this paper and addresses possible future work.

\section{Methods}
\label{sec:method}

In this Section, brief description of MMDenseNet~\cite{takahashi2017multi} is first given, followed by the four invoked or proposed methods, complex ideal ratio mask (cIRM)~\cite{kong2021decoupling}, self-attention, the band-merge-split method inspired by~\cite{luo2023music}, and feature look back, for improving MMDenseNet. While the first three methods aim to improve the separated audio quality, the goal of the feature look back method is to reduce the latency with little degradation of the separated audio quality.

\subsection{The Original MMDenseNet}

\begin{figure}[t]
    \centering
    \includegraphics[width=0.99\columnwidth]{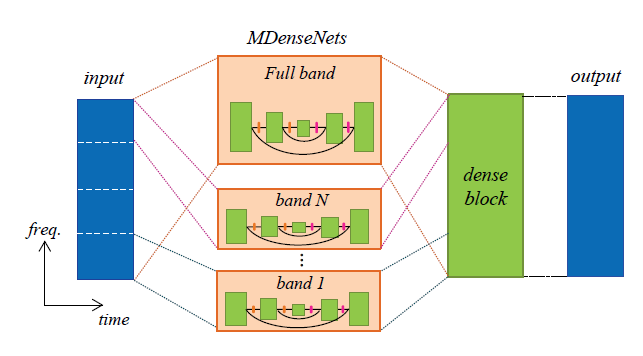}
    \caption{Structure of the original MMDenseNet~\cite{takahashi2017multi}.}
    \label{fig:mmd}
\end{figure}

MMDenseNet~\cite{takahashi2017multi} was proposed by Takahashi and Mitsufuji, which won the signal separation evaluation campaign in 2016~\cite{liutkus20172016} with less number of parameters and faster training speed than other models in the evaluation campaign. The structure of MMDenseNet is shown in~\figref{fig:mmd}. The basic component of MMDenseNet is DenseNet~\cite{huang2017densely} or dense block, which is mainly used to comprise the multi-scale DenseNet (MDenseNet). The MDenseNet is a U-Net like structure, where the encoder part is composed of DenseNets and down-sampling layers, and the decoder part is composed of DenseNets and up-sampling layers. The input spectrogram is splitted into $N$ subbands, and $N$ MDenseNets are used for processing the $N$ subbands, and $1$ MDenseNet is used for processing the full band (i.e. all frequency bands). Finally, $1$ DenseNet is used to combine the outputs of the $N+1$ MDenseNets for obtaining the spectrogram of the separated signal.

\subsection{Complex Ideal Ratio Mask}
\label{subsec:cirm}

While the final results of a source separation task should be spectrograms or signals, the raw output of the neural network in this task can be in different forms, and post-processing techniques can be applied for the output of the network. One of the popular methods is to use the magnitude mask as the output of the separation model, and the output mask is then element-wise multiplied with the input spectrogram to obtain the final output. Besides, for some source separation models, Wiener filter~\cite{wiener1949extrapolation} is used for post-processing to obtain a better final separated results. However, since Wiener filter utilizes all the separated sources for computation, requires a larger amount of computational resources and is unsuitable for this paper (details are shown in Section~\ref{subsec:exp-result}).

\begin{figure}[t]
    \centering
    \includegraphics[width=0.99\columnwidth]{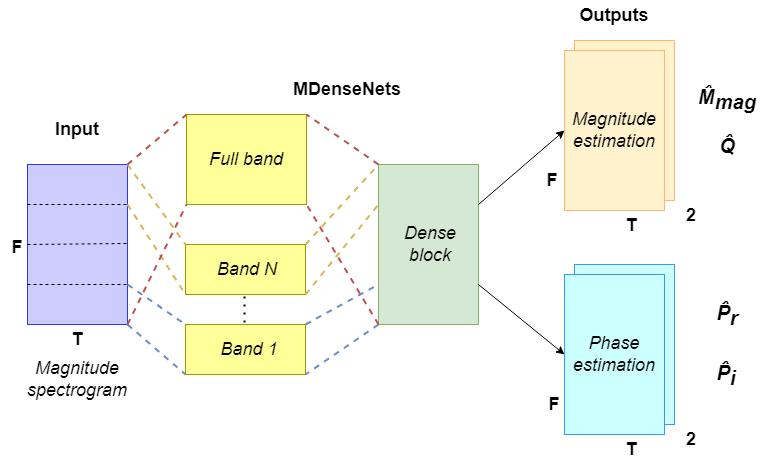}
    \caption{The modified MMDenseNet which uses cIRM as the new output form. $\hat{M}_{mag}$, $\hat{Q}$, $\hat{P}_{r}$ and $\hat{P}_{i}$ are respectively magnitude mask estimation, magnitude estimation, phase estimation of real part, and phase estimation of imaginary part. $F$ and $N$ are respectively 1,025 and 2, and $T$ varies in different experiments in this paper.}
    \label{fig:cIRM}
\end{figure}

In this paper, we investigate the performance of using both magnitude and phase estimation, where their effectiveness has been shown in~\cite{kong2021decoupling}, as the new output form for MMDenseNet. The modified network structure is shown in~\figref{fig:cIRM}, where $\hat{M}_{mag}$, $\hat{Q}$, $\hat{P}_{r}$ and $\hat{P}_{i}$ are respectively magnitude mask estimation, magnitude estimation, phase estimation of real part, and phase estimation of imaginary part. $\hat{M}_{mag}$ and $\hat{Q}$ are used to estimate the magnitude spectrogram, $\hat{P}_{r}$ and $\hat{P}_{i}$ are used to estimate the phase spectrogram, and the final separated result is obtained using estimated magnitude and phase spectrogram and inverse short-time Fourier transform. $F$ and $N$ in~\figref{fig:cIRM} are respectively 1,025 and 2, and $T$ varies in different experiments.

\subsection{Self-Attention}
\label{subsec:sa}

\begin{figure}[t]
    \centering
    \includegraphics[width=0.99\columnwidth]{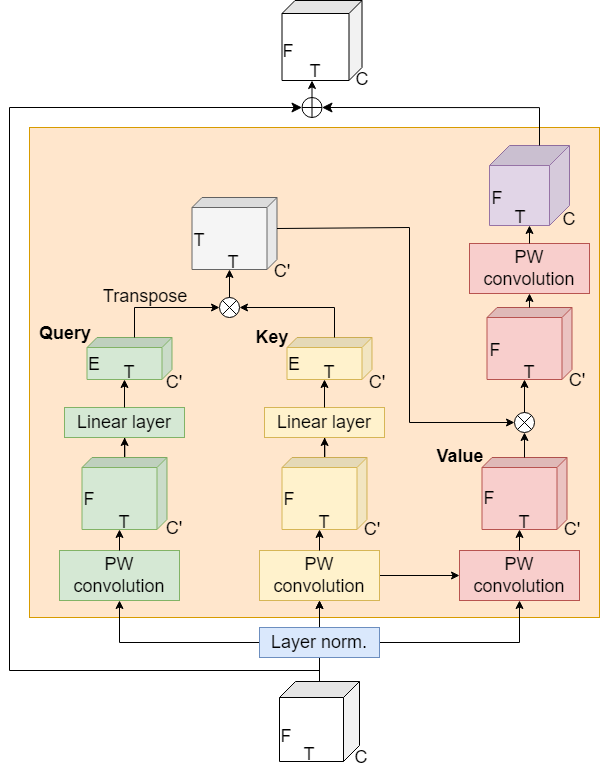}
    \caption{The adjusted structure of the self-attention along time axis. $E$ and $C'$ are respectively set to 20 and 5 in this paper. ``PW'' stands for ``pointwise''.}
    \label{fig:sa_t}
\end{figure}

\begin{figure}[t]
    \centering
    \includegraphics[width=0.99\columnwidth]{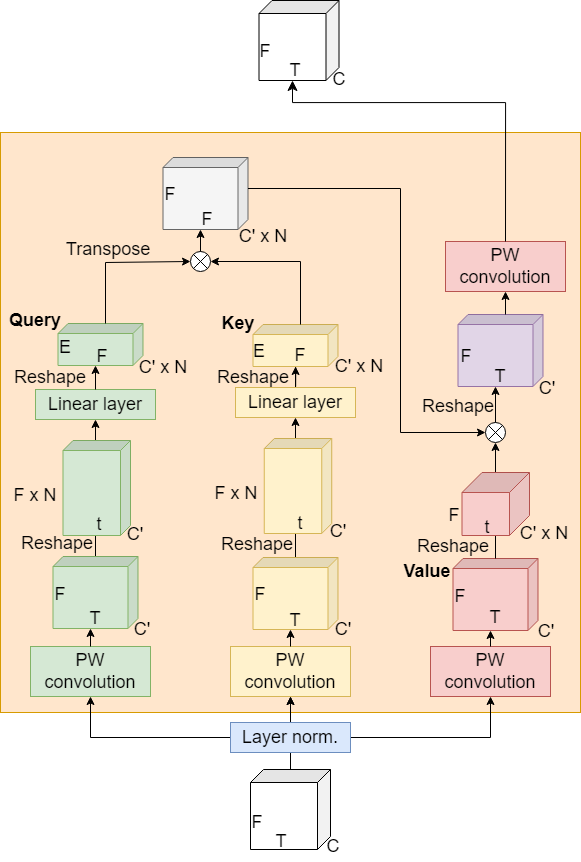}
    \caption{The adjusted structure of the self-attention along frequency axis. $N$ is $T/t$, where $t$ is 16. ``PW'' stands for ``pointwise''.}
    \label{fig:sa_f}
\end{figure}

Due to the great success of using self-attention in various scenarios~\cite{liu2020voice, wang2020axial}, we propose adjusted self-attention structures and apply it to MMDenseNet. The adjusted structure of the self-attention along the time axis is shown in~\figref{fig:sa_t}, where different channels are viewed as different attention heads, and pointwise convolution layers are used to reduce the computational cost by decreasing the number of channels. The adjusted structure of the self-attention along frequency axis is shown in~\figref{fig:sa_f}, which is similar to that of the self-attention along time axis, but chunkwise self-attention is applied, and the residual connection between module input and output is not used. In our setting for chunkwise self-attention, the input spectrogram with length $T$ of the time axis is split into $T/t$ chunks, where $t$ is set to 16 in this paper. Note that our attention structures are similar to that of TF-GridNet~\cite{wang2023tf}, with the following differences: 1) we use layer normalization, 2) normalization is applied only at the beginning of the attention structures, and 3) PReLU is not used.

\begin{figure}[t]
    \centering
    \includegraphics[width=0.99\columnwidth]{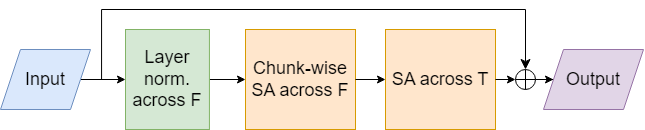}
    \caption{When using both the self-attention along time and frequency axis for fullband MDenseNet, the self-attention along frequency axis is used before the self-attention along time axis. \textbf{SA}: self-attention.}
    \label{fig:fba}
\end{figure}

To apply self attention to the MMDenseNet, we attach the self-attention modules after each dense block of a fullband or subband MDenseNet except the first block. The self-attention along time axis is used for both fullband and subband MDenseNets, but the self-attention along frequency axis is only used for fullband MDenseNet. When using both the self-attention along time and frequency axis for fullband MDenseNet, the self-attention along frequency axis is used before the self-attention along time axis, as shown in~\figref{fig:fba}.

\subsection{Band-Merge-Split Method}
\label{subsec:bs}

\begin{figure}[t]
    \centering
    \includegraphics[width=0.99\columnwidth]{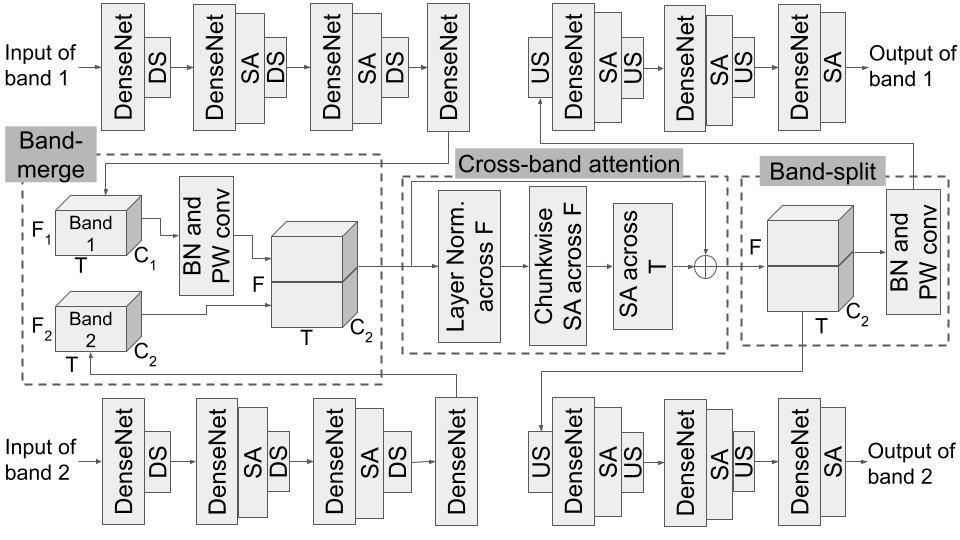}
    \caption{Our band-merge-split method and its connections to the two subband MDenseNets. Skip connections between DenstNets are omitted in this figure. Blocks surrounded by dash lines are respectively band-merge, cross-band attention, and band-split modules.}
    \label{fig:bs}
\end{figure}

The structures of MMDenseNet and band-split RNN~\cite{luo2023music} (BSRNN) are similar in some way since both of them split the input spectrogram into subbands. However, a module similar to the band and sequence modeling module of BSRNN which captures information across time and frequency bands does not exists in MMDenseNet. Therefore, inspired by BSRNN, we use the band-merge-split method for MMDenseNet, and the modified structure is shown in~\figref{fig:bs}.

The band-merge-split method connects two subband MDenseNets and includes three modules: band-merge, cross-band attention, and band-split. The band-merge module concatenates features from the middle of different MDenseNets along the frequency axis. Due to the fact that the features from different MDenseNets have different number of channels, pointwise convolution~\cite{hua2018pointwise} is applied to adjust the number of channels in one of the features before concatenation. The cross-band attention module is used to share information across frequency bands. In order to let each feature be concentrated on its corresponding important frequency bands and time periods, self attention is applied to both the frequency and time axes. The band-split module splits the feature along the frequency axis, and pointwise convolution is applied again to adjust the number of channels back to its original number for further processing.

\subsection{Feature Look Back}
\label{subsec:flb}

Intuitively, source separation models favors longer input and produce low quality separation results when the input duration is short. To maintain the separation quality when the input duration is short, feature look back is used to combine past and current information for obtaining the separation output for the current input. To precisely describe feature look back, following terms should be defined:
\begin{itemize}
  \item Training segment: model input at the training stage, consists of one or multiple training chunks. The size of a training segment is denoted by $N_{\text{tr-s}}$ frames.
  \item Training chunk: model input (consists of multiple frames) at each time at the training stage. The size of a training chunk is denoted by $N_{\text{tr-c}}$ frames. Note that $N_{\text{tr-c}}$ should divide $N_{\text{tr-s}}$.
  \item Training look back chunk: corresponding model input of look back information at each time at the training stage. The size of a training look back chunk is denoted by $N_{\text{tr-lbc}}$ frames.
  \item Test chunk: model input at each time at the test stage. The size of a test segment is denoted by $N_{\text{te-s}}$ frames.
  \item Test look back chunk: corresponding model input of look back information at each time at the test stage. The size of a test look back chunk is denoted by $N_{\text{te-lbc}}$ frames.
\end{itemize}
The latency is reduced by shorten the model input from $N_{\text{tr-s}}$ frames to $N_{\text{tr-c}}$ frames at each time, and the separation quality is possible to be maintained by looking back of past features.

At the training stage, a training segment is split into training chunks, the chunks are feed into the network one by one, and the loss is calculated once all the chunks of a segment are processed by the network. If feature look back is enabled, the hidden representation of the past $N_{\text{tr-lbc}}$ frames are considered together with a current training chunk when processing a current training chunk. The process of the test stage when feature look back is enabled is similar. Test chunks are feed into the network one by one, and past $N_{\text{te-lbc}}$ frames are considered together with a current test chunk when processing a current test chunk.

\begin{figure}[t]
    \centering
    \includegraphics[width=0.99\columnwidth]{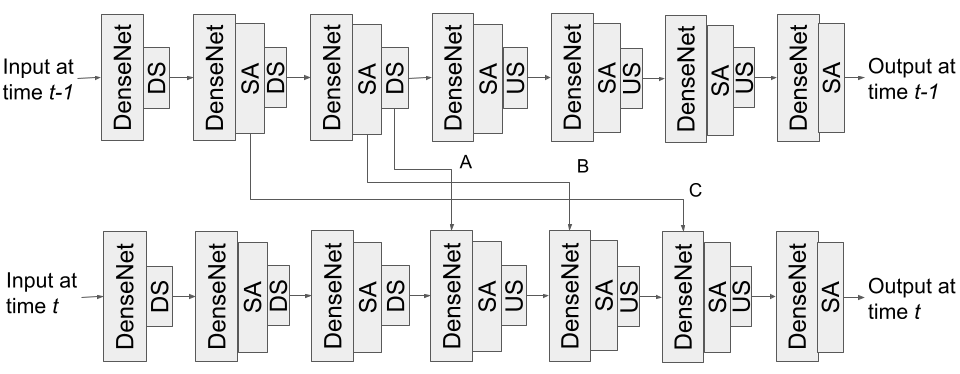}
    \caption{Network structure for each MDenseNet when feature look back is enabled. The input at one timestamp is a training or test chunk. DS, SA, and US are respectively Down Sampling, Self Attention, and Up Sampling. Skip connections between DenstNets are omitted in this figure. A, B, and C are different possible look back connections.}
    \label{fig:looback}
\end{figure}

The network structure for each MDenseNet when feature look back is enabled is shown in~\figref{fig:looback}, where DS, SA, and US are respectively Down Sampling, Self Attention, and Up Sampling, and A, B, and C are different possible look back connections. For simplicity, we only show the case when $N_{\text{tr-c}}$ or $N_{\text{te-c}}$ is equal to $N_{\text{tr-lbc}}$ or $N_{\text{te-lbc}}$ (i.e. look back only for one previous training or test chunk), and skip connections between DenstNets are omitted in the figure. In the experimental Section, we show the results of using only connection A, using both connections A and B (denoted as A+B), and using connections A, B, and C (denoted as A+B+C). Note that our feature look back is different from memory transformers~\cite{burtsev2020memory, zhang2023toward} since the output size is not changed (because the output of the SA blocks after DenseNets are cropped).

\section{Experimental Results}
\label{sec:exp}

\begingroup
\setlength{\tabcolsep}{3pt}
\begin{table*}[ht!]
\begin{center}
\begin{tabular}{lccccrrrrrccc}
\hline
\multirow{2}{*}{Config Name} & \multirow{2}{*}{cIRM}& \multirow{2}{*}{SA} & \multirow{2}{*}{BS} & \multirow{2}{*}{\makecell{Num of\\params (k)}} & \multicolumn{3}{c}{Training} & \multicolumn{2}{c}{Test} & \multirow{2}{*}{SDR} & \multirow{2}{*}{RTF} & \multirow{2}{*}{\makecell{Optimal\\latency (s)}} \\ \cline{6-10}
& & & & & Seg & Chunk & LBC & Chunk & LBC & & & \\
\hline

Raw MMDenseNet & No & No & No & 339 & 256 & 256 & No & 256 & No & 11.162 & 0.368 & 4.38 \\
\makecell[l]{Raw MMDenseNet\\(with Wiener filter)} & No & No & No & 679 & 256 & 256 & No & 256 & No & 13.872 & 0.841 & 9.99 \\
\makecell[l]{Raw MMDenseNet\\(Mag. mask as output)} & No & No & No & 339 & 256 & 256 & No & 256 & No & 13.555 & 0.394 & 4.68 \\

cIRM & Yes & No & No & 343 & 256 & 256 & No & 256 & No & 13.951 & 0.370 & 4.39 \\

\hline

cIRM+SA\_T & Yes & T & No & 574 & 512 & 512 & No & 256 & No & 14.764 & 0.402 & 4.77 \\
cIRM+SA\_T+F & Yes & T+F & No & 578 & 512 & 512 & No & 256 & No & 15.011 & 0.723 & 8.59 \\
cIRM+SA\_T+BS & Yes & T & Yes & 575 & 512 & 512 & No & 256 & No & 14.859 & 0.397 & 4.71 \\

\hline

FLB: A (512/64/64) & Yes & T & No & 574 & 512 & 64 & 64 & 64 & 64 & 14.048 & 0.403 & 1.19 \\
FLB: A (256/32/128) & Yes & T & No & 574 & 256 & 32 & 128 & 32 & 128 & 13.689 & 0.442 & 0.65 \\
FLB: A (128/8/64) & Yes & T & No & 574 & 128 & 8 & 64 & 8 & 64 & 11.322 & 0.692 & 0.25 \\
FLB: A+B & Yes & T & No & 574 & 256 & 32 & 128 & 32 & 128 & 13.538 & 0.468 & 0.69 \\
FLB: A+B+C & Yes & T & No & 574 & 256 & 32 & 128 & 32 & 128 & 13.204 & 0.577 & 0.85 \\
\hline
\end{tabular}
\end{center}
\caption{Separation performance of the accompaniment part for different experimental settings. The unit of segment and chunk size is the number of frames. As a reference, since the sample rate is 44.1 kHz and the window size and hop size are 2048 and 1024 samples, 512, 256, 128, 64, 32, 8 frames are respectively 11.91, 5.97, 3.00, 1.51, 0.77 and 0.21 seconds. The three numbers of ``FLB: A'' are used to denoted different sizes of training segment, training chunk, and look back chunk. \textbf{SA}: self-attention. \textbf{Seg}: segment. \textbf{LBC}: look back chunk. \textbf{FLB}: feature look back.}
\label{tab:result}
\end{table*}
\endgroup

\subsection{Evaluation Metrics}

\begin{figure}[t]
    \centering
    \includegraphics[width=0.9\columnwidth]{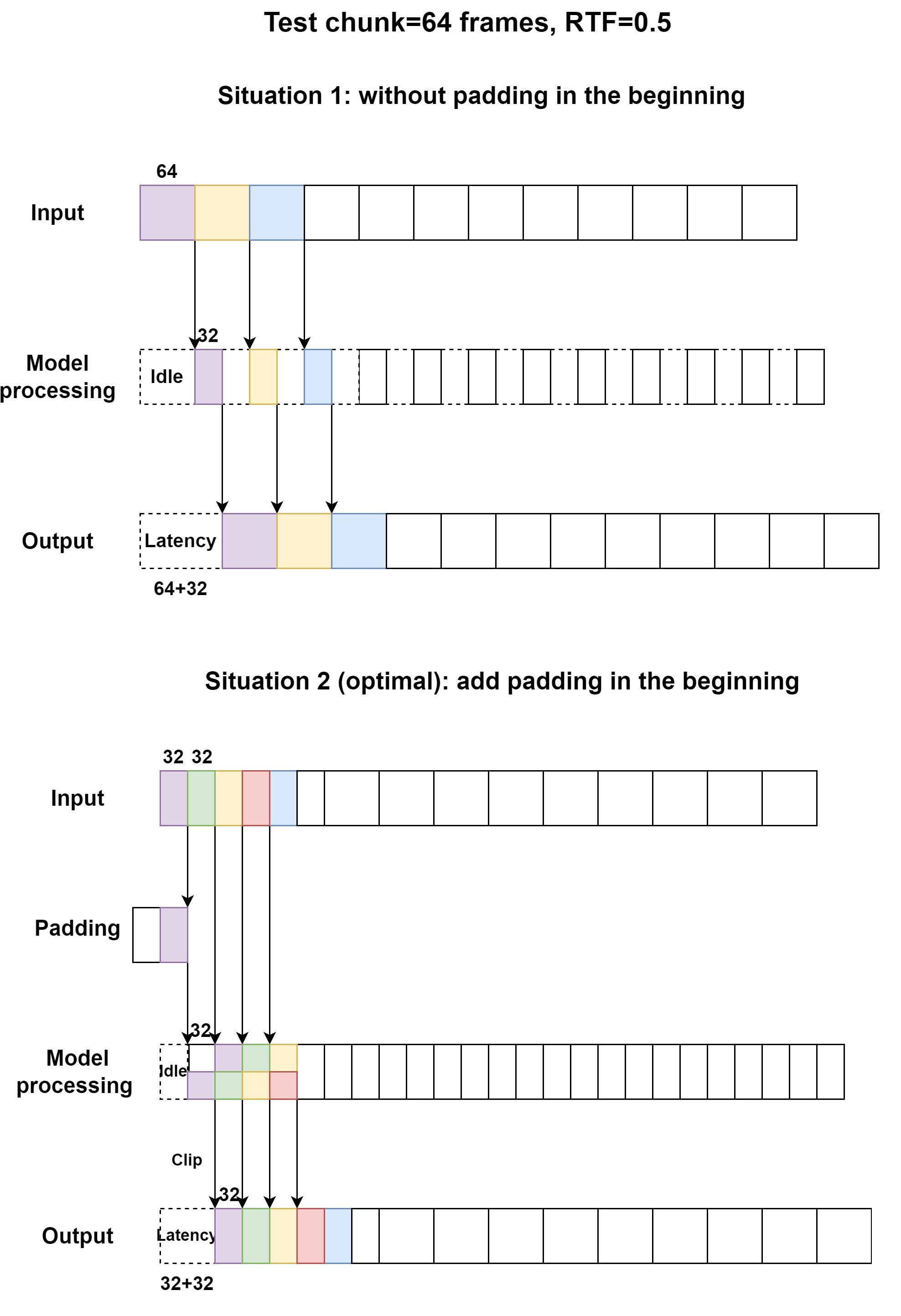}
    \caption{The relationship between the duration of a test chunk, RTF, and latency. Different colors indicate test chunks at different timestamps.}
    \label{fig:rtf_to_lan}
\end{figure}

Source-to-distortion ratio (SDR)~\cite{vincent2006performance}, real-time factor (RTF), and optimal latency are used to evaluate the model performance. SDR is used as the primary evaluation metric in many previous studies~\cite{takahashi2017multi, rouard2023hybrid, mitsufuji2022music}. For a song track, the SDR for each one second segment is first calculated using Museval~\cite{stoter2019museval}, and the song track's SDR is calculated as the median of all one-second segments' SDR. A higher SDR indicates the better separation results. The RTF is defined as the total processing time of a song track divided by the total duration of the song track, and the optimal latency could be therefore calculated as the RTF multiply by the duration of a test chunk multiply by two. The lower RTF and optimal latency indicate that the model runs faster.

The derivation from RTF to optimal latency is shown as follows. Since the case of $RTF \ge 1$ does not meet our application requirement, we consider only the case of $RTF \leq 1$. Generally, latency is the duration of a test chunk ($T$) plus the model processing time ($T + RTF \times T$), as illustrated in the upper part of~\figref{fig:rtf_to_lan}. However, by padding a blank signal of duration $t$ at the beginning and feeding a signal of duration $T$ into the model every $RTF \times T$ time units (where $t \geq RTF \times T$), as depicted in the lower part of~\figref{fig:rtf_to_lan}, the latency can be calculated as follows, and the latency is optimal when the equality holds:

\begin{equation}
\begin{aligned}
Optimal~Latency &= t + RTF \times T \\&\geq RTF \times T + RTF \times T \\&= 2 \times RTF \times T
\label{eq:opt-lan}
\end{aligned}
\end{equation}

\subsection{Dataset and Experimental Setup}

MUSDB18~\cite{musdb18} is used in this study. It contains 150 music tracks of different genres with 44.1 kHz sampling rate. Each track is composed of four channels, including vocal, drum, bass, and the rest of the accompaniment. The goal in this study, accompaniment separation, is to separate the mixture of drum, bass, and the rest of the accompaniment. 86 and 14 tracks are respectively used for training and validation, and 50 tracks are used as the test set for calculating SDR. For RTF calculation, a set of 8 privately collected tracks with an average duration of about 243 seconds are used.

Experiments are conducted in two different machines. The first one is a container running on a machine with Intel(R) Xeon(R) Gold 6154 CPU, occupying 4 CPU cores, 90 GB of host  memory, and a Tesla V100-SXM2 GPU. The second one has an Intel(R) Xeon(R) Silver 4116 CPU, 328 GB of RAM, an NVIDIA Quadro GV100, and an NVIDIA TITAN RTX GPU. For training and test, we use one of the machines simply based on their availability at that time. But for RTF calculation, we run our implementation on the CPU of the later one, and the number of used CPU cores is limited to 1.

The window size and the hop size for short-time Fourier transform are respectively $2,048$ and $1,024$, and the Hann function is used for windowing. The size of the training segment, training chunk, training lookback chunk, test chunk, and test lookback chunk varies in different experiments due to application requirements and will be described in the following subsection. The L1 loss and the ADAM optimizer are used. The initial learning rate is $10^{-3}$ and is multiplied by $0.99$ every $5$ epochs. Due to resource limitations, the batch size is set to $8$ or $16$ for different experiments, and we empirically stop the training process when the validation loss is not significantly decreasing, which is usually no more than $500$ epochs. All models are trained from scratch.

\subsection{Results}
\label{subsec:exp-result}

Separation performance of the accompaniment part for different experimental settings is shown in~\tabref{tab:result}, where the segment and chunk sizes are presented in the number of frames. The first four rows of~\tabref{tab:result} show the settings and evaluation results for the raw MMDenseNet, the MMDenseNet with Wiener filter, the MMDenseNet uses magnitude mask as output, and our improvement using cIRM. After invoking these improvements mentioned in Section~\ref{subsec:cirm}, SDRs are significantly increased (from 11.162 to 13.872, 13.555, or 13.951), and the model using cIRM achieves highest SDR. On the other hand, since our modifications only change the output form (except the one with Wiener filter, which requires larger computational resources), the RTFs remain similar to the raw MMDenseNet. 

By listening to the separated audio of these models, we found that outputs of the MMDenseNet uses magnitude mask as output sounds better than that of the raw MMDenseNet, and outputs of the MMDenseNet with Wiener filter sounds better than that of the MMDenseNet uses magnitude mask as output. Outputs of our improvement using cIRM and the MMDenseNet with Wiener filter sounds similar, but high frequency noise can be perceived in the output of our improvement using cIRM.

The fifth to seventh rows of~\tabref{tab:result} show that both using cIRM with self-attention and band-merge-split methods mentioned in Section~\ref{subsec:sa} and~\ref{subsec:bs} can further improve SDR (from 13.951 to 14.764, 15.011, or 14.859). Using self attention along both time and frequency axes improves SDR the most, but the RTF and latency are also higher, since the chunkwise self attention along the frequency axes requires larger computational resources than the self attention along the time axes. We also notice that the residual vocal sounds less after adding self attention. Two of these configurations, cIRM+SA\_T and cIRM+SA\_T+BS, achieve similar SDRs and RTFs. Despite the fact that the SDR of the cIRM+SA\_T+BS is slightly better than that of cIRM+SA\_T, we choose cIRM+SA\_T for following experiments for the sake of implementation convenience. Besides, we test cIRM+SA\_T using a shorter test chunk of 64 frames, and the SDR decreases from 14.764 to 12.776, showing that we cannot reduce the latency while still maintaining comparable separation quality by only shortening the test chunk. 

Last five rows represent different settings and evaluation results for feature look back mentioned in Section~\ref{subsec:flb}. The three numbers of ``FLB: A'' are used to denoted different sizes of training segment, training chunk, and look back chunk. For the three settings of ``FLB: A'', the results show that when the sizes of training segments and chunks decrease, the SDRs also decrease, verifying out intuition in Section~\ref{subsec:flb}. However, these three SDRs (14.048, 13.689, and 11.322) are still higher than that of the raw MMDenseNet, showing that using feature look back is possible to reduce that latency while maintaining separation quality. Besides, with fixed sizes of training segments and chunks (256 and 32), and as more look back connections are used, the SDRs show a slight downward trend (13.689,  13.538, and  13.204), while the RTFs exhibit a rising trend. These results are caused by the fact that using more look back connections leads to increased computation in the network, and indicate that using shallow information together with deep information may not help increasing the separation quality. Finally, compared to the previous settings, all these FLB settings (except ``FLB: A (128/8/64)'') achieve similar SDRs and listening perspective while significantly decreasing latencies, verifying the effectiveness of our proposed feature look back.

\section{Conclusions and Future Work}
\label{sec:con}

This paper uses the complex ideal ratio mask, self-attention, band-merge-split method, and feature look back to improve MMDenseNet for real-time application. Experimental results show that our method can significantly decrease the RTF and optimal latency while achieving similar SDRs compared to previous study.

Several directions for immediate future work are underway. Currently, the input spectrogram is cut into two subbands, but since the importance of subbands may be different~\cite{takahashi2017multi}, we could cut the input spectrogram into more subbands and investigate different ways of merging them. Moreover, although this study focuses on accompaniment separation due to application requirements, we could also investigate the performance of our model for different source types.



\bibliography{ISMIRtemplate}

\begin{thebibliography}{10}
\providecommand{\url}[1]{#1}
\csname url@samestyle\endcsname
\providecommand{\newblock}{\relax}
\providecommand{\bibinfo}[2]{#2}
\providecommand{\BIBentrySTDinterwordspacing}{\spaceskip=0pt\relax}
\providecommand{\BIBentryALTinterwordstretchfactor}{4}
\providecommand{\BIBentryALTinterwordspacing}{\spaceskip=\fontdimen2\font plus
\BIBentryALTinterwordstretchfactor\fontdimen3\font minus \fontdimen4\font\relax}
\providecommand{\BIBforeignlanguage}[2]{{%
\expandafter\ifx\csname l@#1\endcsname\relax
\typeout{** WARNING: IEEEtran.bst: No hyphenation pattern has been}%
\typeout{** loaded for the language `#1'. Using the pattern for}%
\typeout{** the default language instead.}%
\else
\language=\csname l@#1\endcsname
\fi
#2}}
\providecommand{\BIBdecl}{\relax}
\BIBdecl

\bibitem{rajpura2020effectiveness}
D.~G. Rajpura, J.~Shah, M.~Patel, H.~Malaviya, K.~Phatnani, and H.~A. Patil, ``Effectiveness of transfer learning on singing voice conversion in the presence of background music,'' in \emph{2020 International Conference on Signal Processing and Communications (SPCOM)}.\hskip 1em plus 0.5em minus 0.4em\relax IEEE, 2020, pp. 1--5.

\bibitem{gao2022music}
X.~Gao, C.~Gupta, and H.~Li, ``Music-robust automatic lyrics transcription of polyphonic music,'' in \emph{Proceedings of the 19th Sound and Music Computing Conference}, 2022, pp. 325--332.

\bibitem{liu2020voice}
Y.~Liu, B.~Thoshkahna, A.~Milani, and T.~Kristjansson, ``Voice and accompaniment separation in music using self-attention convolutional neural network,'' \emph{arXiv preprint arXiv:2003.08954}, 2020.

\bibitem{rouard2023hybrid}
S.~Rouard, F.~Massa, and A.~D{\'e}fossez, ``Hybrid transformers for music source separation,'' in \emph{ICASSP 2023-2023 IEEE International Conference on Acoustics, Speech and Signal Processing (ICASSP)}.\hskip 1em plus 0.5em minus 0.4em\relax IEEE, 2023, pp. 1--5.

\bibitem{takahashi2017multi}
N.~Takahashi and Y.~Mitsufuji, ``{Multi-scale Multi-band Densenets for Audio Source Separation},'' in \emph{2017 IEEE Workshop on Applications of Signal Processing to Audio and Acoustics (WASPAA)}.\hskip 1em plus 0.5em minus 0.4em\relax IEEE, 2017, pp. 21--25.

\bibitem{kong2021decoupling}
Q.~Kong, Y.~Cao, H.~Liu, K.~Choi, and Y.~Wang, ``Decoupling magnitude and phase estimation with deep resunet for music source separation,'' \emph{arXiv preprint arXiv:2109.05418}, 2021.

\bibitem{luo2023music}
Y.~Luo and J.~Yu, ``{Music Source Separation With Band-Split RNN},'' \emph{IEEE/ACM Transactions on Audio, Speech, and Language Processing}, 2023.

\bibitem{jansson2017singing}
A.~Jansson, E.~Humphrey, N.~Montecchio, R.~Bittner, A.~Kumar, and T.~Weyde, ``Singing voice separation with deep u-net convolutional networks,'' 2017.

\bibitem{stoter2019open}
F.-R. St{\"o}ter, S.~Uhlich, A.~Liutkus, and Y.~Mitsufuji, ``Open-unmix-a reference implementation for music source separation,'' \emph{Journal of Open Source Software}, vol.~4, no.~41, p. 1667, 2019.

\bibitem{kim2021kuielab}
M.~Kim, W.~Choi, J.~Chung, D.~Lee, and S.~Jung, ``Kuielab-mdx-net: A two-stream neural network for music demixing,'' \emph{arXiv preprint arXiv:2111.12203}, 2021.

\bibitem{liutkus20172016}
A.~Liutkus, F.-R. St{\"o}ter, Z.~Rafii, D.~Kitamura, B.~Rivet, N.~Ito, N.~Ono, and J.~Fontecave, ``The 2016 signal separation evaluation campaign,'' in \emph{Latent Variable Analysis and Signal Separation: 13th International Conference, LVA/ICA 2017, Grenoble, France, February 21-23, 2017, Proceedings 13}.\hskip 1em plus 0.5em minus 0.4em\relax Springer, 2017, pp. 323--332.

\bibitem{huang2017densely}
G.~Huang, Z.~Liu, L.~Van Der~Maaten, and K.~Q. Weinberger, ``Densely connected convolutional networks,'' in \emph{Proceedings of the IEEE conference on computer vision and pattern recognition}, 2017, pp. 4700--4708.

\bibitem{wiener1949extrapolation}
N.~Wiener, \emph{Extrapolation, interpolation, and smoothing of stationary time series: with engineering applications}.\hskip 1em plus 0.5em minus 0.4em\relax The MIT press, 1949.

\bibitem{wang2020axial}
H.~Wang, Y.~Zhu, B.~Green, H.~Adam, A.~Yuille, and L.-C. Chen, ``Axial-deeplab: Stand-alone axial-attention for panoptic segmentation,'' in \emph{European conference on computer vision}.\hskip 1em plus 0.5em minus 0.4em\relax Springer, 2020, pp. 108--126.

\bibitem{wang2023tf}
Z.-Q. Wang, S.~Cornell, S.~Choi, Y.~Lee, B.-Y. Kim, and S.~Watanabe, ``Tf-gridnet: Making time-frequency domain models great again for monaural speaker separation,'' in \emph{ICASSP 2023-2023 IEEE International Conference on Acoustics, Speech and Signal Processing (ICASSP)}.\hskip 1em plus 0.5em minus 0.4em\relax IEEE, 2023, pp. 1--5.

\bibitem{hua2018pointwise}
B.-S. Hua, M.-K. Tran, and S.-K. Yeung, ``Pointwise convolutional neural networks,'' in \emph{Proceedings of the IEEE conference on computer vision and pattern recognition}, 2018, pp. 984--993.

\bibitem{burtsev2020memory}
M.~S. Burtsev, Y.~Kuratov, A.~Peganov, and G.~V. Sapunov, ``Memory transformer,'' \emph{arXiv preprint arXiv:2006.11527}, 2020.

\bibitem{zhang2023toward}
W.~Zhang, K.~Saijo, Z.-Q. Wang, S.~Watanabe, and Y.~Qian, ``Toward universal speech enhancement for diverse input conditions,'' in \emph{2023 IEEE Automatic Speech Recognition and Understanding Workshop (ASRU)}.\hskip 1em plus 0.5em minus 0.4em\relax IEEE, 2023, pp. 1--6.

\bibitem{vincent2006performance}
E.~Vincent, R.~Gribonval, and C.~F{\'e}votte, ``Performance measurement in blind audio source separation,'' \emph{IEEE transactions on audio, speech, and language processing}, vol.~14, no.~4, pp. 1462--1469, 2006.

\bibitem{mitsufuji2022music}
Y.~Mitsufuji, G.~Fabbro, S.~Uhlich, F.-R. St{\"o}ter, A.~D{\'e}fossez, M.~Kim, W.~Choi, C.-Y. Yu, and K.-W. Cheuk, ``Music demixing challenge 2021,'' \emph{Frontiers in Signal Processing}, vol.~1, p. 808395, 2022.

\bibitem{stoter2019museval}
F.~St{\"o}ter and A.~Liutkus, ``museval 0.3. 0,'' 2019.

\bibitem{musdb18}
\BIBentryALTinterwordspacing
Z.~Rafii, A.~Liutkus, F.-R. St{\"o}ter, S.~I. Mimilakis, and R.~Bittner, ``The {MUSDB18} corpus for music separation,'' Dec. 2017. [Online]. Available: \url{https://doi.org/10.5281/zenodo.1117372}
\BIBentrySTDinterwordspacing

\end{thebibliography}

\end{document}